\documentclass[a4paper,usenatbib,twocolumn]{mnras}

\usepackage[T1]{fontenc}
\usepackage{ae,aecompl}

\usepackage{graphicx}	
\usepackage{amsmath}	
\usepackage{amssymb}	

\title[Shot noise and multiple tracers]{Shot noise in multi-tracer constraints on $f_\text{NL}$ and relativistic projections: Power Spectrum.}

\author[D. Ginzburg et al.]{
Dimitry Ginzburg$^{1}$\thanks{E-mail: dima.ginzburg@gmail.com} and
Vincent Desjacques$^{1,2}$
\\
$^{1}$Physics department, Technion, Haifa 3200003, Israel\\
$^{2}$Asher Space Science Institute, Technion, Haifa 3200003, Israel
}

\date{Accepted 2020 April 20. Received 2020 March 10; in original form 2019 December 01}

\pubyear{2020}

\hypersetup{draft}

\begin{document}
\label{firstpage}
\pagerange{\pageref{firstpage}--\pageref{lastpage}}
\maketitle

\begin{abstract}
Multiple tracers of the same surveyed volume can enhance the signal-to-noise on a measurement of local primordial non-Gaussianity and the relativistic projections. Increasing the number of tracers comparably increases the number of shot noise terms required to describe the stochasticity of the data. Although the shot noise is white on large scales, it is desirable to investigate the extent to which it can degrade constraints on the parameters of interest. 
In a multi-tracer analysis of the power spectrum, a marginalization over shot noise does not degrade the constraints on $f_\text{NL}$ by more than $\sim 30$\% so long as halos of mass $M\lesssim 10^{12}M_\odot$ are resolved.
However, ignoring cross shot noise terms induces large systematics on a measurement of $f_\text{NL}$ at redshift $z<1$ when small mass halos are resolved. These effects are less severe for the relativistic projections, especially for the dipole term.
In the case of a low and high mass tracer, the optimal sample division maximizes the signal-to-noise on $f_\text{NL}$ and the projection effects simultaneously, reducing the errors to the level of $\sim 10$ consecutive mass bins of equal number density. 
We also emphasize that the non-Poissonian noise corrections that arise from small-scale clustering effects cannot be measured with random dilutions of the data. Therefore, they must either be properly modeled or marginalized over.
\end{abstract}




\section{Introduction}
\label{sec:introduction}

Fluctuations in the galaxy number counts can be expressed as a series expansion in the matter density and tidal shear. The coefficients of this series are the galaxy bias parameters \citep[][for a recent review]{Desjacques:2016bnm}. They are intimately related to the underlying halo mass function and to the (nonlinear) gravitational evolution of density perturbations \citep[see, e.g.,][]{kaiser:1984,bbks,cole/kaiser:1989,Mo:1995cs,catelan/etal:1999,sheth/tormen:1999}.
Furthermore, one must take into account projection effects resulting from the propagation of light between the observed galaxies and the detector. One of these projection effects is the well-known "Kaiser effect", which arises from the relative line of sight velocity between the galaxy and the observer \citep{kaiser:1987}. Other contributions arise from gravitational lensing, wide-angle effects etc. 
\citep[see][for early work in the context of galaxy clustering]{hamilton/culhane:1996,moessner/etal:1998,matsubara:2000,hui/etal:2007}.
As shown in \cite{2009PhRvD..80h3514Y} \citep[see also][]{2010PhRvD..82h3508Y,lewis/challinor:2011,bonvin/durrer:2011,Baldauf:2011bh,jeong/schmidt/hirata:2012}, all these effects can be consistently accounted for in a relativistic formulation of galaxy counts. These effects become more significant as the survey volume increases (and are thus expected to be relevant for upcoming surveys such as Euclid or LSST).
Finally, one must include shot noise contributions to account for the discreteness of the galaxy distribution. These are usually taken to be Poissonian, but data (galaxies) and numerical simulations (halos) have shown evidence for non-Poissonian effects \citep{CasasMiranda:2002on,Hamaus:2010im,Paech:2016hod}.

Statistical measures of galaxy number counts, which encode important information on viable cosmologies, suffer from sampling variance and shot noise.
Sampling variance can be mitigated by combining different tracers of the same surveyed volume \citep{2009PhRvL.102b1302S}. Such a multi-tracer approach can enhance the signal-to-noise on a power spectrum measurement of local primordial non-Gaussianity \citep{mcdonald/seljak:2009,2011PhRvD..84h3509H,Raccanelli:2014awa,alonso/ferreira:2015}. Similar gains could in principle be achieved on, e.g., the logarithmic growth rate of structures \citep{white/etal:2009,Hamaus:2012ap}.
These Fisher matrix analyzes have been - at least partly - validated with the analysis of N-body simulations, in which the galaxy positions are approximated by the center-of-mass locations of dark matter halos \citep{2011PhRvD..84h3509H}. Moreover, a suitable (mass) weighting can be applied to mitigate the shot noise as well, and further reduce the uncertainty on the amount of local primordial non-Gaussianity \citep{Seljak:2009af,slosar:2009,cai/etal:2011,dePutter:2014lna}.

Although the non-Gaussian bias and relativistic projections share a similar scale-dependence, this degeneracy can be broken with multiple tracers of the large scale structure \citep{Yoo:2012se,camera/etal:2015}. In fact, the relativistic projections become detectable only when multiple tracers are used \citep{Yoo:2012se,2015PhRvD..92f3525A,2015ApJ...812L..22F,2019ApJ...872..126M,2019ApJ...870L...4M,2020MNRAS.492.1513G}. In light of current attempts to measure these effects and, in particular, constrain the amount of local primordial non-Gaussianity (a detection of which would have profound implications), it is important to gauge the different sources of systematics in order to maximize the signal-to-noise. The latter is not expected to be much larger than a few $\sigma$'s for $|f_\text{NL}|\sim 1$. In this paper, we will focus on the shot noise. Although it is expected to be weakly degenerate with the local non-Gaussianity signal and the relativistic projections (owing to their specific scale-dependence), it is prudent to check the extent to which uncertainties on the parameters of interest - obtained from a multi-tracer analysis - degrade when we marginalize over the multiple shot noise terms, are biased by a Poissonian noise assumption, or can be reduced with an optimal division of the survey data into samples.

\begin{figure}
	\includegraphics[width=0.45\textwidth]{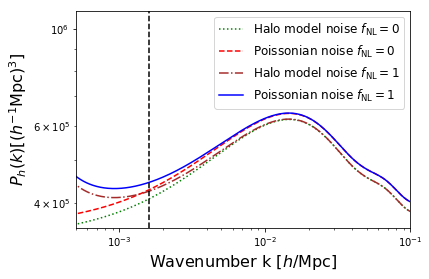}
	\caption{Power spectrum of $z=1$ halos with mass in the range $10^{14}-10^{17}[h^{-1} M_{\odot}]$. Only the non-Gaussian bias from $\delta_\text{Proj}$ is included. The various curves are computed as labelled accordingly. The dashed vertical line represents the minimum wavenumber $k$ measured from a surveyed volume of $60 \, h^{-1} \rm{Gpc^3}$.}
	\label{fig:PS1} 
\end{figure}


To characterize the shot noise, we will rely on the halo model, which has been shown to provide a reasonable fit to simulation data \citep{Hamaus:2010im,2011PhRvD..84h3509H,2017PhRvD..96h3528G}. The shot noise model, along with our linear theory description of the observed large scale galaxy power spectrum, are briefly reviewed in \S\ref{sec:model}. Our results are presented in details in \S\ref{sec:results}, and summarized in \S\ref{sec:conclusions}. 


\section{Model}
\label{sec:model}

We will closely follow the notation of \cite{Yoo:2012se} to facilitate comparison with this work. 

\subsection{Large scale clustering}

For simplicity, we assume throughout that galaxies and halos are interchangeable, that is, each dark matter halo hosts exactly one galaxy located at its center. HODs could be used to make this correspondence more realistic. In any case, this simplification does not affect our conclusions. 
Since we are interested in the large scale properties of the galaxy distribution, we retain the linear contribution to the observed galaxy density only. In Fourier space, this becomes
\begin{equation}
\delta_h(k) = b_1 \delta_m(k) + \delta_\text{Proj}(k)+\epsilon(k)  \;.
\end{equation}
Here, $\delta_h(k)$ and $\delta_m(k)$ are the Fourier modes of the (real space) halo and dark matter overdensity, $\epsilon(k)$ is the shot noise, and $b_1$ is the linear bias. At low $k$, the shot noise asymptotes to a white noise contribution $\varepsilon\equiv\epsilon(k\to 0)$. We adopt this approximation throughout the paper.
Furthermore, to avoid clutter, we use $b$ instead of $b_1$ to designate the linear bias (so that we can label the linear bias of different samples with a subscript). This will not add any confusion since we do not consider contributions beyond linear theory. 

We consider the possibility of a local primordial non-Gaussianity parametrized by $f_\text{NL}$, in which case the linear, scale-independent bias $b$ becomes \citep{2008PhRvD..77l3514D,2008JCAP...08..031S,2008ApJ...677L..77M}
\begin{equation}
b\rightarrow b(k) \equiv b+f_{\textrm{NL}}\frac{\partial\ln\bar n}{\partial\ln\sigma_8}\frac{3\Omega_{m0} H_0^2}{k^2 T(k) D(z) c^2} \;.
\end{equation}
We will approximate the $\sigma_8$-derivative by $\delta_c(b_1-1)$, where $\delta_c\simeq 1.686$ is the linear critical overdensity for a spherical collapse, but one should bear in mind this is a bad approximation for certain halo definitions \citep{Biagetti:2016ywx}.
Furthermore, $T(k)$ is the matter transfer function normalized to unity as $ k\rightarrow 0$, $D(z)$ is the linear growth rate normalized to $1+z$ and $\Omega_{m0}$ is the present-day matter density. Unless otherwise stated, we adopt the fiducial value of $f_\text{NL}=0$ in all illustrations.

In the plane-parallel approximation and for narrow redshift bins, the linear projection terms ranked according to their power of $\mathcal{H}/k$ are given by 
\begin{equation}
\label{eq:deltaproj}
\delta_\textrm{Proj}(k)=\bigg[\frac{\mathcal{P}}{(k/\mathcal{H})^2}-i\mu_k\frac{\mathcal{R}}{(k/\mathcal{H})}+f\mu_k^2\bigg]\delta_m(k) \;,
\end{equation} 
where $\mu_k$ is the cosine angle between the line-of-sight direction and the wave vector, $\mathcal{H}$ is the conformal Hubble rate, and $f$ is the logarithmic growth rate $f={d\ln\delta_m}/{d\ln a}$.
Ignoring the gravitational lensing and the integrated Sachs-Wolfe contributions, the functions $\mathcal{P}$ and $\mathcal{R}$ take the form \citep{Yoo:2012se}
\begin{align}
\mathcal{P}&=ef-\frac{3}{2}\Omega_m(z) \bigg[ e+f-\frac{1+z}{H}\frac{dH}{dz}+(5p-2)(2-\frac{1}{\mathcal{H}r}) \bigg] \nonumber\\
\mathcal{R}&=f\bigg[e-\frac{1+z}{H}\frac{dH}{dz}+(5p-2)(1-\frac{1}{\mathcal{H}r})\bigg]
\end{align}
where $e$ is the time evolution of the mean galaxy number density and $p$ is the luminosity function slope of source galaxy population at the threshold, 
\begin{equation}
p=\frac{\partial\log\bar{n}}{\partial \mathcal{M}}\;,\quad e=\frac{\partial\ln\bar{n}}{\partial\ln(1+z)} \;.
\end{equation} 
Here, $\bar{n}_g$ is the average tracer (galaxy) density as a function of observed redshift $z$, while $\mathcal{M}$ is the absolute magnitude. For simplicity, we will set $e=3$ (i.e. constant comoving number density as in \cite{Yoo:2012se}) throughout, although our survey specifications do not correspond to narrow redshift intervals. 

When Eq.(\ref{eq:deltaproj}) is generalized to the case of multiple tracers, $p$ will likely vary among the tracers. For sake of comparison with \cite{Yoo:2012se} however, we will assume a common value of $p=0$ or $p=0.4$ among the tracers. Setting $p=0$ is appropriate to surveys with very low luminosity threshold, whereas $p=0.4$ signifies that one is considering a flux limited survey dominated by the galaxies at the peak of the luminosity function.
We do not consider higher values of $p$ (corresponding to a survey in which the high luminosity tail solely is sampled) since, in this case, low mass halos are not resolved and, therefore, the signal-to-noise for the projection effects and the local primordial non Gaussianity is expected to be low (this is exemplified in e.g. Fig.\ref{fig:HP1}). This emphasizes the fact that, in reality, $p$ will likely depend on the minimum halo mass $M_\text{min}$ resolved in the survey.

Note also that the dipole term $\propto \mu_k$ leads to an imaginary contribution to the cross-power spectrum of two different tracers \citep{mcdonald:2009}, which makes $\mathcal{R}$ weakly degenerate with the other model parameters \citep[see][for a first attempt to measure this effect]{gaztanaga/etal:2017}.

\subsection{White noise}

We take into account non-Poissonian contributions to the white noise piece of the halo power spectrum as in \citep{Hamaus:2010im}. Non-Poissonian white noise is a generic outome of (nonlinear) biased processes \citep{Durrer:2002zu,BeltranJimenez:2010bb}.
At large halo mass, the physical origin of these corrections is small scale exclusion of halo centers \citep{Mo:1995cs,Sheth:1998xe,Smith:2006ne,Baldauf:2013hka}, which makes the low-$k$ white noise contribution to the halo power spectrum sub-Poissonian.

The halo model approach advocated in \cite{Hamaus:2010im} provides a reasonable fit to the numerical results across a wide range of masses \citep[see also][for an extension to the bispectrum]{2017PhRvD..96h3528G} . In this model, the elements $\mathcal{E}_{ij}$ of the shot noise matrix take the form
\begin{equation}
\label{eq:HMnoise}
    \mathcal{E}_{ij} =\frac{1}{\bar n}\delta_{ij}^\text{K} - \frac{\big(b_i\overline{{M}}_j+b_j\overline{{M}}_i\big)}{\bar\rho_\text{m}}+b_i b_j \frac{\langle n M^2\rangle}{\bar\rho_\text{m}^2}
\end{equation}
where, here and henceforth, $\bar n$ denotes an integrated number density whereas $n=n(M)$ designates the differential mean number counts $dn/dM$. Indices $i,j$ designate the mass bins, $\overline{{M}}_i$ is the average halo mass of bin $i$, $\bar\rho_\text{m}$ is the average comoving matter density, and the brackets $\langle ... \rangle$ represent an average over all halos. 

The second and third term in Eq.~(\ref{eq:HMnoise}) arise from absorbing the 1-halo contributions to the halo-matter and matter-matter power spectrum into the shot noise. They are the sole contribution to the off-diagonal elements of $\mathcal{E}_{ij}$, which are generically non-vanishing in contrast to Poissonian noise. At high halo mass, the second term dominates over the third, and makes the noise sub-Poissonian. At small halo mass, the third becomes larger and, thus, the noise is super-Poissonian.

As shown in Fig.\ref{fig:PS1}, the white noise flattens the galaxy power spectrum at low $k$. This effect is somewhat degenerate with that of the non-Gaussian bias.

\begin{table}
        \begin{center}
	\begin{tabular}{|l|l|l|}
		\hline\hline
		& $z$   & $V (h^{-1} \rm{Gpc^3})$   \\ \hline
		Survey 1 & 0.5 & 59  ($z=0\sim1$) \\ \hline
		Survey 2 & 2   & 410 ($z=1\sim3$) \\ \hline\hline
	\end{tabular}
		\caption{The parameters for the two survey configurations used in all illustrations. The volumes are calculated from the lower and upper redshift limit given in brackets. Similar survey characteristics are considered in  \protect\cite{Yoo:2012se}.}
		\label{tab:table1}
       \end{center}
\end{table}

\subsection{Fisher matrix}

Since deviations from Gaussianity are actually small (except, possibly, for the modes on the largest scales), the likelihood for a measurement of the Fourier modes $\boldsymbol{\delta^{\textrm{obs}}}=(\delta_{g1},\dots,\delta_{gN})^\top$ of $N$ different galaxy tracers at a single wavenumber $\textbf{\textit{k}}$ is
\begin{equation}
\mathcal{L}=\frac{1}{(2\pi)^{(N/2)}(\det \textbf{C})^{(1/2)}} \exp \bigg[ -\frac{1}{2}\boldsymbol{\delta^{\textrm{obs}\dagger}}\textbf{C}^{-1} \boldsymbol{\delta^{\textrm{obs}}}\bigg]
\end{equation}
The corresponding covariance matrix is $\textbf{C}=\langle \boldsymbol{\delta^{\textrm{obs}}} \boldsymbol{\delta^{\textrm{obs}\dagger}} \rangle=\boldsymbol{bb^{\dagger}}P_m+\boldsymbol{\mathcal{E}}$, $\boldsymbol{\mathcal{E}}=\langle \boldsymbol{\varepsilon} \boldsymbol{\varepsilon}^T \rangle$ and $P_m$ is the matter power spectrum in the comoving gauge. The Fisher matrix for the set of model parameters $\theta_a$ is
\begin{equation}
\label{eq:FisherSingleMode1}
F_{ab}=\bigg \langle -\frac{\partial^2\ln\mathcal{L}}{\partial \theta_a \partial \theta_b} \bigg \rangle= \frac{1}{2} \textrm{Tr}\bigg[\textbf{C}^{-1} \textbf{C}_a \textbf{C}^{-1} \textbf{C}_b\bigg] \;,
\end{equation}
where the indices $a$ and $b$ run over the model parameters, $\textbf{C}_a=\partial \textbf{C}/\partial \theta_a$.
For the parameters  $f_\text{NL}$ ,$\mathcal{R}$, $\mathcal{P}$ , assuming the shot noise term does not depend on any of the parameters above, the Fisher matrix can be written in the following way:
\begin{equation}
\label{eq:FisherSingleMode}
F_{ab}=\frac{\alpha}{1+\alpha}\gamma_{ab}+\frac{\mathrm{Re}(\beta_a\beta_b-\alpha \beta_a\beta_b^*)}{(1+\alpha )^2} \;,
\end{equation}
where we have defined ${\alpha=\boldsymbol{b}^\dagger\boldsymbol{\mathcal{E}}^{-1}\boldsymbol{b}P_m}$, $\beta_a=\boldsymbol{b}^{\dagger} \boldsymbol{\mathcal{E}}^{-1}\boldsymbol{b}_a P_m$, $\gamma_{ab}=\mathrm{Re}(\boldsymbol{b}_a^{\dagger}\boldsymbol{\mathcal{E}}^{-1}\boldsymbol{b}_b)P_m$
and $\boldsymbol{b_a}=\partial \boldsymbol{b}/\partial \theta_a$. \citep{2011PhRvD..84h3509H,Yoo:2012se}

To calculate the functions $\alpha$, $\beta_a$ and $\gamma_{ab}$, we use the Sheth-Tormen formula for the linear bias \citep{sheth/tormen:1999} and a $\Lambda$CDM power spectrum consistent with Cosmic Microwave Background (CMB) constraints \citep[from the WMAP7 analysis more specifically, see][]{2011ApJS..192...18K}. 

The Fisher matrix of the survey is given by the integral of Eq.~(\ref{eq:FisherSingleMode}) across the relevant range of wavenumbers. In what follows, we will include Fourier modes between $k_\text{min}=2\pi/V^{1/3}$ (the fundamental mode) and  $k_\text{max}=0.03\,h {\rm Mpc}^{-1}$. This rather low values of $k_\text{max}$ ensures that the linear approximation adopted here is valid throughout. 

As emphasized above, we will assume there is a one-to-one correspondence between the surveyed galaxies and the host dark matter halos and, moreover, the clustering properties of galaxies are fully determined by the host halo mass $M$ (with a maximum halo mass of $M=10^{17}\, h^{-1} M_\odot$ in all practical calculations).
We will, by default, construct galaxy samples such that they share the same comoving density. We will relax this assumption later. Furthermore, to facilitate the comparison with previous studies -- in particular, \cite{Yoo:2012se} -- we consider various surveys with mean survey and volume as listed in Table~\ref{tab:table1}.

\begin{figure*}
	\includegraphics[width=1\textwidth]{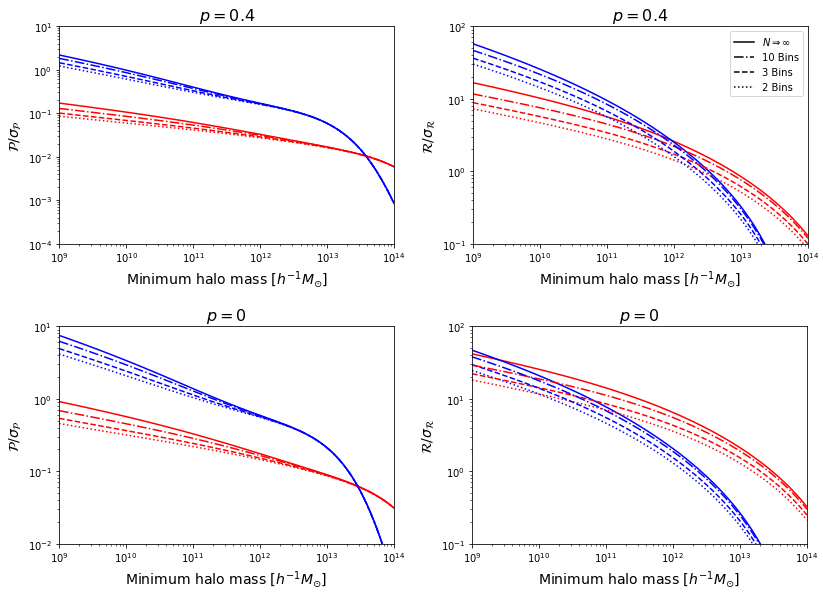}
	\caption{Dependence of the signal-to-noise $\mathcal{P}/\sigma_\mathcal{P}$ and $\mathcal{R}/\sigma_\mathcal{R}$ on the number of tracers. The red and blue curves show the results for survey 1 and 2, respectively. They agree with the findings of \protect\cite{Yoo:2012se}.}
	\label{fig:HP1}
\end{figure*}

\section{Results}
\label{sec:results}

\subsection{Model parameters and preliminary checks}
\label{sec:preliminarytests}

We focus on the detectability of $\mathcal{R}$, $\mathcal{P}$, $f_\text{NL}$ in the galaxy power spectrum, assuming that all the other parameters (i.e. $b_i$, $f$ etc.), except the noise terms, have been constrained using, for instance, a combination of power spectra and bispectra measurements at wavenumber larger than $k_\text{max}$. Since $p$ assumes a unique value among all tracers, our model parameters thus are $(\mathcal{R},\mathcal{P},f_\textrm{NL},\epsilon_{ij})$, where $\epsilon_{ij}$ represents the $N(N+1)/2$ white noise contributions to the auto- and cross-power spectra of the $N$ galaxy tracers. Our aim is understand the influence of uncertainties in the shot noise estimates on a measurement of $\mathcal{R}$, $\mathcal{P}$ and $f_{\textrm{NL}}$. 
In principle, cosmological parameters (such as $n_s$, $\sigma_8$, $\Omega_{\Lambda}$ e.t.c) and bias parameters are estimated from the survey as well, they should be marginalized over as well. However, in \cite{PhysRevD.95.123513}  it was concluded that the effect is typically less then 20-30 percent for the multitracer case for the Fisher matrix (and so even less for the error) and therefore we ignore it. Note, however, that though the effect is small, it is not negligible and should be taken into account in estimation of the errors in a real survey.

To begin, we performed consistency checks with the literature, and assessed the extent to which the marginalized errors on $\mathcal{R}$, $\mathcal{P}$, $f_\text{NL}$ depend on the shot noise model. We considered the halo and Poisson models outlined above and found there is almost no difference between the two models (regardless of the number of bins), in agreement with the findings of \cite{PhysRevD.95.123513} (who focused on $f_\text{NL}$ though). 

\subsection{Varying the number of tracers}

Next, we checked the dependence of the measurement significance on the number of bins. In a real survey, only a finite number $N$ of bins can be considered, while the idealized (continuous) limit $N\to\infty$ provides a lower bound to the uncertainties. Fig.~\ref{fig:HP1} shows the signal-to-noise $\mathcal{P}/\sigma_\mathcal{P}$ and $\mathcal{R}/\sigma_\mathcal{R}$ as a function of $N=2$, 3, 10 and $N\to\infty$. The marginalized errors $\sigma_\mathcal{P}$ and $\sigma_\mathcal{R}$ were obtained upon fixing all the other model parameters to their fiducial value.

Increasing the number of bins from 2 to 10 can increase the signal-to-noise by roughly a factor of 2 to 3, especially if low mass halos $M\lesssim 10^{12}\,h^{-1} M_\odot$ are included in the survey. For $\mathcal{P}$, splitting the data into $N\geq 3$ tracers improves the signal-to-noise only if the minimum halo mass is below $10^{12}\,h^{-1}M_\odot$. In any case, the signal-to-noise remains too low to allow a detection of this term at redshift $z<1$.

\begin{figure*}
	\includegraphics[width=1\textwidth]{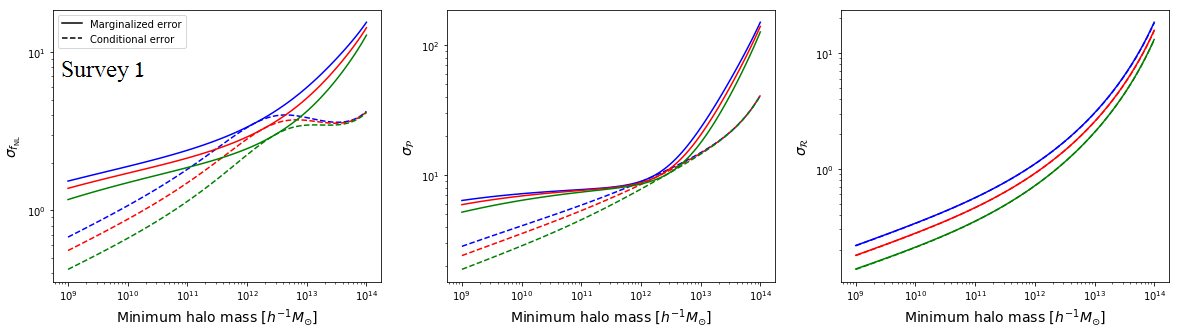}
	\includegraphics[width=1\textwidth]{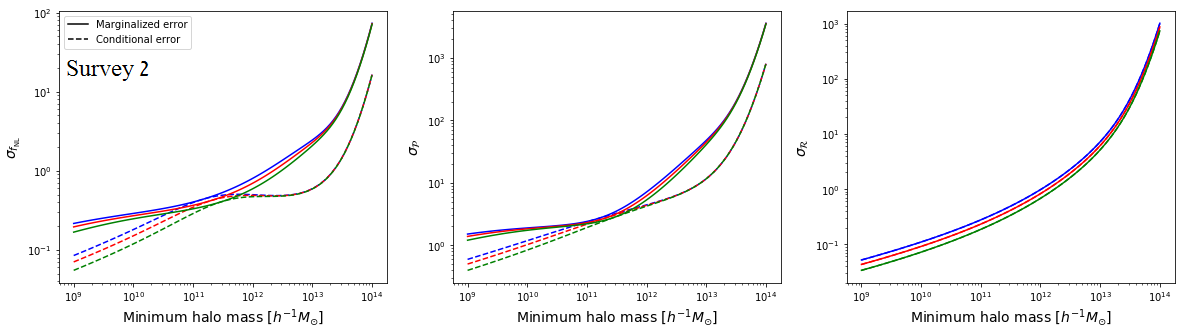}
	\caption{Marginalized (solid) and conditional (dashed) errors for survey 1 (top panels) and survey 2 (bottom panels) for different number of tracers with equal number density. The blue, red and green curves represent the 2, 3 and 10 tracers case. All the model parameters $(\mathcal{R},\mathcal{P},f_\text{NL},\epsilon_{ij})$ are allowed to vary.}
	\label{fig:HP2}
\end{figure*}

Allowing $f_\text{NL}$ and the shot noise terms to vary changes the errors as shown in Fig.~\ref{fig:HP2}. Solid and dashed curves indicate the marginalized and conditional (i.e. all the remaining parameters are fixed) for $N=2$, 3 and 10 bins. A value of $p=0$ is assumed, but we checked that the results remain qualitatively the same when $p=0.4$ is adopted instead. 
The peculiar dipolar dependence of the $\mathcal{R}$ contribution translates into very similar conditional and marginalized errors.
By contrast, these errors differ significantly at halo mass $M\lesssim 10^{12}\,h^{-1}M_\odot$ for $\mathcal{P}$ and $f_\text{NL}$. 

\begin{figure*}
	\includegraphics[width=1\textwidth]{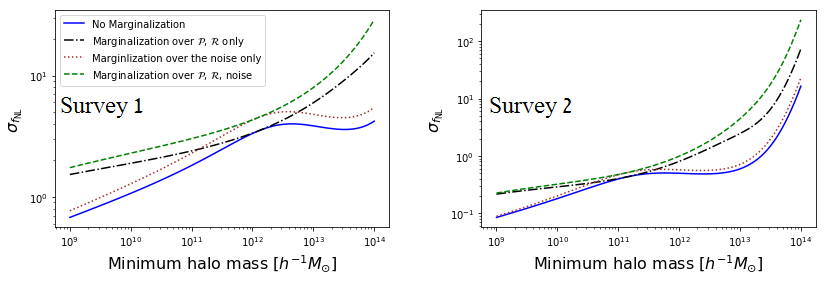}
	\caption{Marginalized vs. conditional error on $f_\text{NL}$ for two tracers of equal number density, and $p=0$. Results are shown for the two survey specifications considered here.}
	\label{fig:HP3}
\end{figure*}

Since the non-Gaussian bias and the relativistic projections are inversely proportional to powers of $k$, marginalizing over the white noise will not significantly expand the errors.
We have performed the same test for both 2 and 3 tracers of equal number density (in the latter case, the marginalization is over 6 independent noise parameters) and found very similar results. Therefore, only the 2 tracers case is shown in Fig.~\ref{fig:HP3} for illustration.  
A marginalization over the diagonal and off-diagonal shot noise terms degrade the errors by 20\% (resp. 10\%) for $M_\text{min}=10^{10}\,h^{-1}M_\odot$ and 30\% (resp. 15\%) for $M_\text{min}=10^{12}\,h^{-1}M_\odot$ for survey 1 (resp. survey 2).
For three tracers, our numerical checks return comparable error degradations. We thus speculate that the numbers quoted here are representative of a generic $N\geq 2$ multi-tracer power spectrum analysis, that is, marginalizing over all the shot noise terms does not expand the error on $f_\text{NL}$ by more than $\sim 30$\%. When $M_\text{min}\lesssim 10^{11}h^{-1}M_\odot$, most of the error degradation arises from the GR projection effects. For $\mathcal{P}$, the error degradation is somewhat smaller at same value of $M_\text{min}<10^{12} \,h^{-1}M_\odot$.

\subsection{Optimizing the bin width}

Until now, we assumed that the mass bins share the same number density. When $N$ becomes very large, this restriction does not matter anymore because the sum over the mass bins uniformly converges towards the ideal case $N\to\infty$ (in which sums can be replaced by integrals). For a small number of bins however, the division in bins of equal number density is, of course, suboptimal (the optimal division minimizes the conditional error and, as we shall see later, the marginalzed error as well). Let us illustrate this in the case of two tracers.

Let $M^*\in [M_\text{min},M_\text{max}]$ define the boundary of the two mass bins. To find the optimal value of $M^*$ which minimizes the error on $f_\text{NL}$, consider the Fisher matrix element 
\begin{equation}
F_{f_{\text{NL}} f_{\text{NL}}}=\frac{\alpha}{1+\alpha}\gamma_{f_{\text{NL}} f_{\text{NL}}}+\frac{\text{Re}(\beta_{f_{\text{NL}}}\beta_{f_{\text{NL}}}+\alpha \beta_{f_{\text{NL}}} \beta_{f_{\text{NL}}} ^*) }{(1+\alpha)^2} \;.
\end{equation} 
Under the assumption that the imaginary part of $b$ arising from the relativistic dipole term is much smaller than the real part, the linear bias $b$ is real and we get:
\begin{align}
F_{f_{\text{NL}} f_{\text{NL}}}=\frac{\alpha}{1+\alpha}\gamma_{f_{\text{NL}} f_{\text{NL}}}+\frac{(1-\alpha)\beta_{f_{\text{NL}}}^2}{(1+\alpha)^2}\;.
\end{align} 
The low and high mass bins can be characterized by their effective linear (Gaussian) bias $b_1$ and $b_2$; and their number densities $\bar n_1$ and $\bar n_2$ defined as
\begin{gather}
b_1=\frac{1}{\bar n_1}\int_{M_{\text{min}}}^{M^*}\!\!dM\,n(M) b(M)\nonumber \\ b_2=\frac{1}{\bar n_2}\int_{M^*}^{M_{\text{max}}}\!\!dM\, n(M) b(M)\nonumber \\
\bar n_1=\int_{M_{\text{min}}}^{M^*}\!\!dM\, n(M)\;,\quad
\bar n_2=\int_{M_{*}}^{M_{\text{max}}}\!\!dM\, n(M) \nonumber \\ \bar n_{12}\equiv \bar n_1+\bar n_2 \;.
\end{gather}
Since the halo model and the Poissonian approximation to the shot noise return very similar measurement errors (see \S\ref{sec:preliminarytests} for a discussion), we assume Poissonian noise to simplify the calculation. For two tracers, this implies
\begin{equation}
\boldsymbol{\mathcal{E}^{-1}}=
\left(\begin{matrix}
\bar n_1 & 0 \\
0 & \bar n_2 
\end{matrix} \right) \;.
\end{equation}
Therefore, we find
\begin{align}
\beta_{f_{\text{NL}}} &=K\textbf{\textit{b}}^\dagger\boldsymbol{\mathcal{E}^{-1}}(\textbf{\textit{b}}-\mathbf{1}) P_m \nonumber \\
&=\alpha K - (b_1 \bar n_1 + b_2 \bar n_2) P_m  
\end{align}
and
\begin{align}
\gamma_{f_{\text{NL}} f_{\text{NL}}} &= \left[ K^2 ( \textbf{\textit{b}}-\mathbf{1})^\dagger \boldsymbol{\mathcal{E}^{-1}} (\textbf{\textit{b}}-\mathbf{1}) \right] P_m \nonumber \\
&= K^2 \alpha - 2K^2 P_m (b_1 \bar n_1 + b_2 \bar n_2) + K^2 P_m \bar n_{12} \;,
\end{align}
where
\begin{equation}
K=\frac{3\delta_{cr}\Omega_{m0}H_0^2}{k^2 T(k)D_{md}(\tau)} 
\end{equation}
is the amplitude of the non-Gaussian bias. 
Here, $\delta_c$ is the spherical collapse threshold, $D_{md}(\tau)$ is the linear growth normalized to $a(\tau)$ during the matter-dominated epoch, $k$ is the wavenumber, $T(k)$ is the transfer function (normalized to unity in the limit $k\to 0$), $\Omega_{m0}$ is the matter density parameter today and $H_0$ is the Hubble constant today. 

\begin{figure*}
	\includegraphics[width=1\textwidth]{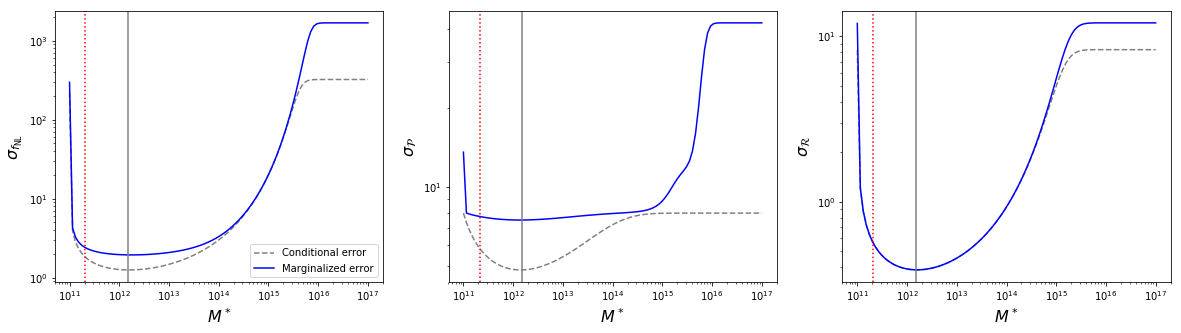}
	\caption{Conditional (dotted) and marginalized (solid) errors on $f_{\text{NL}}$, $\mathcal{R}$ and $\mathcal{P}$ as a function of the halo mass $M^*$ used two split the halo population into a low and high mass sample. The gray vertical line indicate the value of $M^*$ which simultaneously minimaizes the error on $f_\text{NL}$, $\mathcal{P}$ and $\mathcal{R}$ (see text). The red vertical line shows the value of the border when the bins have even halo density. For illustration purposes, results are shown for a minimum and maximum halo mass of $10^{11}$ and $10^{17}$ $h^{-1} M_{\odot}]$, respectively. A low redshift survey with $p=0$ and $f_{\textrm{NL}}=0$ is also assumed.}
	\label{fig:Uneven}
\end{figure*}

The optimum $M^*$ is reached when $F_{f_{\text{NL}} f_{\text{NL}}}$ is maximized. The extremum condition 
$\frac{\partial}{\partial M^*} F_{f_{\text{NL}} f_{\text{NL}}}(M^*)=0$ is satisfied only when 
\begin{equation}
\label{eq:AlphaCondition}
\frac{\partial \alpha(M^*)}{\partial M^*}=0 \;.
\end{equation}
This is due to the fact that, since $M_\text{min}$ and $M_\text{max}$ are held fixed,
\begin{equation}
\frac{\partial (b_1 \bar n_1 + b_2 \bar n_2)}{\partial M^*}=0
\end{equation}
and, thus, 
\begin{equation}
\frac{\partial \gamma_{f_{\text{NL}} f_{\text{NL}}}}{\partial M^*}=K^2 \frac{\partial \alpha}{\partial M^*}\quad \mbox{and}\quad \frac{\partial \beta_{f_{\text{NL}}}}{\partial M^*}=K\frac{\partial \alpha}{\partial M^*} \;.
\end{equation}
Using the relations
\begin{align}
\frac{\partial b_1}{\partial M^*}&=\frac{n(M^*)}{\bar n_1}\big(-b_1+b(M^*)\big) \nonumber \\ \frac{\partial b_2}{\partial M^*}&=\frac{n(M^*)}{\bar n_2}\big(b_2-b(M^*)\big) \nonumber \\
\frac{\partial \bar n_1}{\partial M^*}&=n(M^*) \nonumber \\
\frac{\partial \bar n_2}{\partial M^*} &=-n(M^*) \;,
\end{align}
the condition
\begin{align}
0=\frac{\partial \alpha(M^*)}{\partial M^*}=\frac{\partial (b_1^2 \bar n_1 + b_2^2 \bar n_2)}{\partial M^*} P_m
\end{align}
implies
\begin{equation}
n(M^*) (b_2-b_1) \big[b_1 + b_2 - 2 b(M^*)\big]P_m=0\;.
\end{equation}
Since $b(M)$ is a monotonically increasing function, $b_1$ can not be equal to $b_2$. Hence, we must have
\begin{equation}
\label{eq:MassCondition}
b(M_*)=\frac{b_1+b_2}{2}\;.
\end{equation}
Since this relations is true for any $k$, it is also true when the Fisher matrix elements are integrated over $k$ space.
The same relation Eq.(\ref{eq:MassCondition}) remains valid for $\mathcal{R}$ and $\mathcal{P}$.
To show this, define 
\begin{equation}
K_1 = \frac{1}{(k/\mathcal{H})^2}\;, \quad K_2=i \mu \frac{1}{(k/\mathcal{H})} \;.
\end{equation}
Therefore, 
\begin{align}
F_{\mathcal{RR}}&=\frac{\alpha}{1+\alpha}K_2^2 \bar n_{12} P_m + \frac{-K_2^2}{1+\alpha}   \nonumber \\
F_{\mathcal{PP}}&=\frac{\alpha}{1+\alpha}K_1^2 \bar n_{12} P_m + \frac{K_1^2 (1-\alpha)}{(1+\alpha)^2}
\end{align}
Clearly, $\frac{\partial F_{\mathcal{RR}}(M^*)}{\partial M^*}=0$ and  $\frac{\partial F_{\mathcal{PP}}(M^*)}{\partial M^*}=0$ when $M^*$ also satisfies $\frac{\partial \alpha}{\partial M^*}=0$. Note that Eq.(\ref{eq:MassCondition}) is consistent with the optimal weighting functions proposed by \cite{slosar:2009} (which are proportional to the linear bias).

It is straightforward to show that the condition Eq.(\ref{eq:MassCondition}) corresponds to a minimum of $\alpha(M^*)$. This reflects the fact that the information is maximized when the sampling variance $1+\alpha$ is minimized or, alternatively, when the FKP weight \citep{FKP} $(1+\alpha)^{-1}$ is maximized. For a single tracer, the sampling variance is minimized for a zero bias tracer \citep{castorina/seljak/etal:2018}

To illustrate this point further, Fig.~\ref{fig:Uneven} compares the equal number density vs. optimal division for $N=2$ bins. Solid and dashed curves show the marginalized and conditional errors for $f_{\text{NL}}$, $\mathcal{R}$ and $\mathcal{P}$ when $M^*$ is varied from $M_\text{min}=10^{11}h^{-1}M_{\odot}$ until $M_\text{max}=10^{17}h^{-1}M_{\odot}$.
Scenario 1, with the fiducial $f_{\mathrm{NL}}=0$, is adopted for illustration. For the parameters  $f_\text{NL}$ ,$\mathcal{R}$, $\mathcal{P}$, the minimum for the conditional error is also the minimum for the marginalized error, and it corresponds to the value of $M^*$ which realizes the condition Eq.~(\ref{eq:MassCondition}). This is due to the fact that all the relevant components of the Fisher matrix reach an extremum when $\frac{\partial \alpha(M^*)}{\partial M^*}=0$ (Since all the contribution to $\gamma_{ij}$ and $\beta_i$ are either linear in $\alpha$ or have zero derivative with respect to $M^*$). Therefore, the relevant components of the inverse Fisher matrix will also reach an extremum for the same value of $M^*$.
The division shown as the red dotted line, for which $\bar n_1=\bar n_2$, is clearly suboptimal \citep[up to a factor of 2, in agreement with the findings of][]{slosar:2009}.
Our conclusions are insensitive to the value of $M_\text{min}$, $p$ or the survey details. 

\begin{figure*}
	\includegraphics[width=1\textwidth]{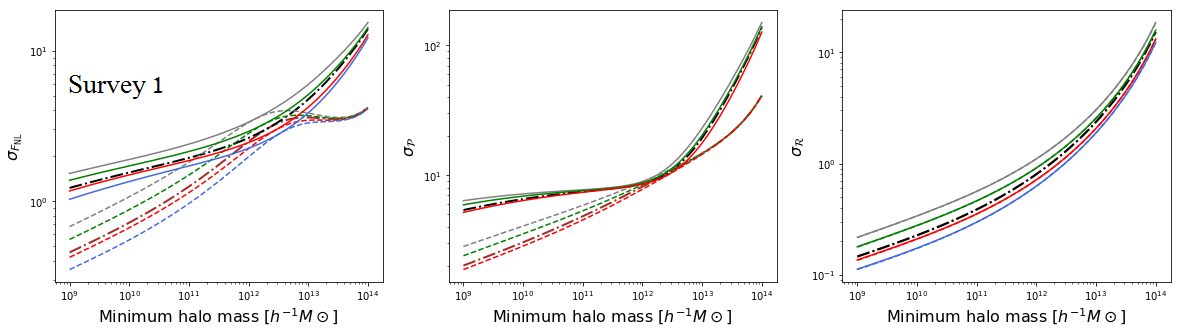}
	\includegraphics[width=1\textwidth]{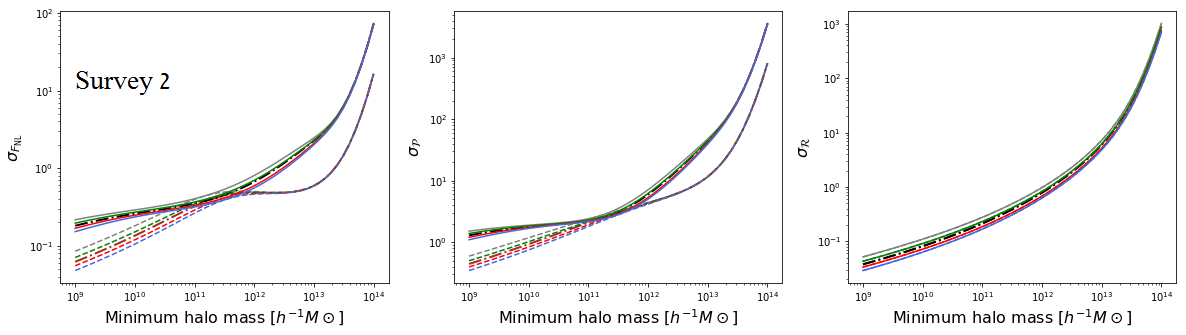}
	\caption{Conditional and marginalized errors (solid and dashed curves, respectively) as a function of the minimum halo mass when the number of tracers is 2 (gray curves), 3 (green curves), 10 (red curves) and 100 (blue curves). The samples are constructed such that they share the same number density. For the two tracer case however, we also constructed the 2 samples such that they maximize the FKP weight (see text). The corresponding errors are shown as the dotted-dashed curves (brown for conditional and black for marginalized). Results are shown for scenario 1 and 2 with the fiducial value $f_{\textrm{NL}}$=0, and with $p=0$.}
	\label{fig:UnevenBins}
\end{figure*}

In figure Fig.~\ref{fig:UnevenBins} we show uncertainties on $f_{\text{NL}}, \mathcal{R},\mathcal{P}$ as a function of minimum halo mass bin for the two survey characterizations and assuming $p$=0 (the results for $p$=0.4 are very similar) for different numbers of tracers sharing the same number density (color curves as indicated in the caption). For the two tracers case, we also constructed the bins such that the FKP weight is maximized as discussed above. The resulting "optimized" errors are shown as the dotted-dashed curves. They are comparable to those obtained with 10 tracers of equal number density. 

\subsection{Systematic errors}

\begin{figure*}
	\includegraphics[width=1\textwidth]{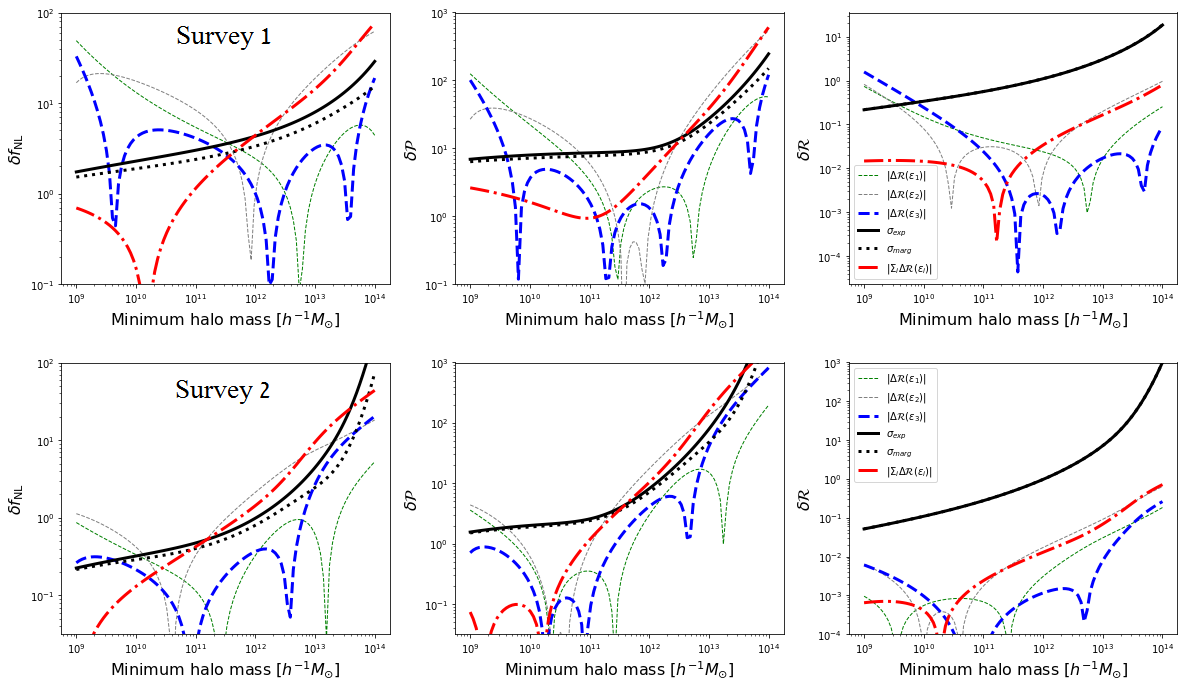}
	\caption{Systematic errors on $f_{\textrm{NL}}$, $\mathcal{P}$ and $\mathcal{R}$ caused by the Poissonian assumption in a two tracer analysis. The dotted black curve represents the statistical error marginalized over all parameters but the noise, while the solid black curve shows the uncertainty when all the parameters - including the noise - are marginalized over. The blue dashed curve is the systematic error arising from the off-diagonal noise term solely, while the green and gray dotted curves indicate the systematic error induced by the diagonal noise terms. The red, dashed-dotted line is the combined effect of all of them.}
	\label{fig:sys}
\end{figure*}

The Fisher Matrix formalism can also be applied to estimate the systematic errors caused by incorrect model assumptions arising, for instance, from a wrong value assignment to a model parameter \citep{2009arXiv0906.0664H,Heavens_2007}. In the present case, we are interested in calculating the systematic bias on $f_{\text{NL}}$, $\mathcal{R}$ and $\mathcal{P}$ due to a misestimation of the noise. Following \cite{2009arXiv0906.0664H}, let $M'$ be a model with $n'$ parameters nested in a bigger model $M$ with $p$ additional parameters. Furthermore, assume that the $p$ parameters are fixed to some incorrect values in $M'$, and that $M'$ is used to analyze the data. One can easily show that the best-fit values of the $n'$ parameters shift from their true value by
\begin{equation}
{\delta\theta'_{\alpha}=-(\textbf{F}'^{-1})_{\alpha\beta}{G}_{\beta\zeta}\delta\psi_{\zeta}}
\end{equation}
where $\theta_{\alpha}$ are the $n'$ common parameters, whereas $\psi_{\zeta}$ represent the $p$ additional parameters in $M$, i.e. $\alpha,\beta=1..n',\zeta=1..p$. Furthermore,
\begin{equation}
{G_{\beta\zeta}=\frac{1}{2}\mathrm{Tr}[\textbf{C}^{-1}\textbf{C},_\beta \textbf{C}^{-1} \textbf{C},_{\zeta}+\textbf{C}^{-1}(\mu,_{\zeta}\mu^T,_{\beta}+\mu,_{\beta}\mu^T,_{\zeta})]}
\end{equation}
where $\mu$ is the ensemble average over the data values and  $x,_{\alpha}=\partial x/\partial \theta_{\alpha}$.  
Since $\langle\delta\rangle=0$ by definition and  ${\psi_{\zeta}=\mathcal{E}}$ we get 
\begin{align}
&{G_{\beta\varepsilon}=\frac{1}{2}\mathrm{Tr}[\textbf{C}^{-1}\textbf{C},_\beta \textbf{C}^{-1} \textbf{C},_{\varepsilon}]}\\ \nonumber
&{\delta\theta'_{\alpha}=-(\textbf{F}'^{-1})_{\alpha\beta}G_{\beta\zeta}\delta{\mathcal{E}}_{\zeta}}
\end{align}
where  $\alpha,\beta=1..n'$.
In the case of a single tracer, $\boldsymbol{\mathcal{E}}$ is a scalar whereas, for $m$ tracers, it is a vector of dimension $p=m(m+1)/2$. Systematic errors will arise when e.g. $\boldsymbol{\mathcal{E}}$ is fixed to the Poissonian expectation while the actual noise is non-Poissonian. To quantify this effect, we consider two tracers and set $\delta\boldsymbol{\mathcal{E}}$ to be the difference between the Poissonian and halo model predictions. In Fig.~\ref{fig:sys}, we show the systematic errors $\delta f_\text{NL}$, $\delta\mathcal{P}$ and $\delta\mathcal{R}$ caused by a shift in one of the shot noise terms solely, and by the sum of all three. The systematic error arising from the off-diagonal term $\epsilon_{12}$ - which is often ignored in multi-tracer studies - becomes significant for low redshift $f_\text{NL}$ constraints as soon as the mass drops below $\sim 10^{12}h^{-1}M_{\odot}$. Therefore, it is essential to include this term in the theoretical model, and marginalize over it in the absence of a reliable theoretical prediction. For the high redshift survey, the systematic error on $f_\text{NL}$ and $\mathcal{P}$ is mainly driven by the shot noise of the low mass sample, $\epsilon_{22}$. Note that $\mathcal{R}$ is generally immune to any systematics, unless the low redshift survey probes halos of mass $M\lesssim 10^{10}h^{-1}M_{\odot}$. In this case however, a marginalization would not degrade the constraints.

\subsection{Sensitivity to the minimum wavenumber}

There is the possibility that some of the modes are severely contaminated by foregrounds, especially on large scales, and must be discarded. To account for this possibility, we performed similar checks in which the smallest $k_\text{min}$ used for the calculation of the Fisher information was taken to be an even integer multiple of the fundamental mode. 

Since the number of modes decrease, the errors increase, albeit not in a similar way. Namely, while the statistical errors do not expand significantly for $\mathcal{R}$ (especially for the high redshift survey) for $k_\text{min}$ as large as 6 times its fiducial value, they increase noticeably for $\mathcal{P}$ and $f_\text{NL}$. This is consistent with the fact that $\mathcal{P}$ and $f_\text{NL}$ terms are proportional to $1/k^2$, while the $\mathcal{R}$ term is proportional to $1/k$. As a result, taking out small $k$ modes affects  $\mathcal{P}$ and $f_\text{NL}$ terms more strongly. The systematic errors caused by wrong assumptions about the shot noise (as discussed above) change (in absolute magnitude) by a comparable amount.

Reducing $k_\text{max}$ (i.e. taking half the fiducial value) affects the uncertainties in the following way. For $\mathcal{R}$, this leads to a slight increase in the statistical error while there is no discernable difference for the systematic shift. By contrast, the statistical error on $\mathcal{P}$ and $f_\text{NL}$ slightly increases while the corresponding systematic error appears to show the opposite trend. Overall, varying either $k_\text{max}$ or $k_\text{min}$ around their fiducial value does not significantly affect our previous results, which are therefore robust to the exact range of wavenumbers considered for the analysis \citep[see also the discussion in][]{PhysRevD.95.123513}. However, the degradation in the constraints could change noticeably if a much larger $k_\text{max}$ (i.e., $\gtrsim 0.1\,h {\rm Mpc}^{-1}$) is used. However, the linear approximation made in this paper would not be valid anymore. 

\subsection{Estimating the shot noise from the data}

Can we extract the corrections to Poisson noise directly from the data? Consider the method outlined in \cite{Andoetal:2018} for instance. Namely, split a given halo catalog $\delta_h$ with linear bias $b$ into two random subsamples $\delta_{h,1}$ and $\delta_{h,2}$, with bias $b_1=b_2=b$ and number density $(2/\bar n)$. Then, form the half-difference (HD) map \citep[in the notation of][]{Andoetal:2018}
\begin{equation}
    \text{HD} \equiv \frac{1}{2}\left(\delta_{h,1}-\delta_{h,2}\right)\;.
\end{equation}
Since the two subsamples are equally clustered, the power spectrum of HD reads
\begin{equation}
\langle(\text{HD})^2\rangle= \frac{1}{4} \Big[\langle \epsilon_1^2\rangle+\langle\epsilon_2^2\rangle-2\langle\epsilon_1\epsilon_2\rangle\Big]\;,
\end{equation}
where angle brackets denote an ensemble average. Eq.~(\ref{eq:HMnoise}) then implies that the non-Poissonian contributions all cancel out, and one is left with
\begin{equation}
   \langle(\text{HD})^2\rangle = \frac{1}{\bar n}
\end{equation}
because the two halo subsamples do not overlap. This result remains valid for actual galaxy samples, and is independent of the details of the shot noise model. It shows that the non-Poissonian corrections to the power spectrum which truly arise from the small-scale clustering properties of large scale structure tracers \citep[i.e., small-scale exclusion effects, see][]{Smith:2006ne,Baldauf:2013hka} cannot be measured using random dilution of the parent sample. However, non-Poissonian effects in the galaxy distribution unrelated to clustering could be measured with such an approach.

\section{Conclusions}
\label{sec:conclusions}

We investigated the impact of shot noise uncertainties arising for multi-tracer constraints on local primordial non-Gaussianity (parametrized by $f_\text{NL}$) and line-of-sight projection effects (parametrized by $\mathcal{P}$ and $\mathcal{R}$). We have considered a low ($\bar z=0.5$) and high-redshift ($\bar z=2$) survey in which galaxies are idealized as halo centers of mass. We do not expect our conclusions to change significantly for more realistic halo occupation distributions. All the errors are computed at the Fisher matrix level. Based on the survey configurations considered here, our conclusions can be summarized as follows:
\begin{itemize}
    \item In a multi-tracer analysis of the power spectrum, a marginalization over all the shot noise terms does not degrade the constraints on $f_\text{NL}$ by more than $\sim 30$\% regardless of the number of tracers, so long as halos of mass $M\lesssim 10^{12}h^{-1}M_\odot$ are resolved.
    \item The uncertainties on $f_\text{NL}$, $\mathcal{P}$ and $\mathcal{R}$ can be simultaneously optimized upon maximizing the FKP weight. As a result, two optimized tracers give errors comparable to those obtained with ten tracers of equal number density.
    \item Ignoring the off-diagonal shot noise contributions induces large systematics on a measurement of $f_\text{NL}$ at redshift $z<1$. At high redshift, they lead to significant biases only if halos down to masses $\lesssim 10^{10}h^{-1}M_{\odot}$ are surveyed.
    \item If off-diagonal shot noise terms are ignored, then it is better to use the Poissonian approximation when low mass halos $M\lesssim 10^{11}\,h^{-1}M_\odot$ are resolved.
    \item Constraints on $\mathcal{R}$, which parametrizes the amplitude of the dipole term, are immune to most of these effects, but one must include the off-diagonal shot noise terms especially if low mass halos $M\lesssim 10^{10}h^{-1}M_\odot$ are surveyed.
    \item It is not possible to measure the non-Poissonian noise corrections from random dilutions of the data, because they actually arise from small-scale clustering effects (projected onto large scale through Fourier transformation).
\end{itemize}
Strictly speaking, these results are valid so long as the model parameters do not vary significantly across the surveyed volume. A tomographic decomposition can be performed when this condition is not fullfilled. 
Furthermore, we have assumed throughout that the linear bias of the surveyed data is well-constrained from, e.g., an independent measurement of the bispectrum. In principle, the multi-tracer approach can be extended to higher-order statistics and, in particular, combined measurements of the power spectrum and bispectrum as considered in \cite{SPHEREx,yamauchi/yokoyama/etal:2017}. Our discussion can be readily extended to this case using predictions from the halo model \citep[see for instance][]{2017PhRvD..96h3528G}.

\section*{Acknowledgements}

D.M. would like to thank Eiichiro Komatsu for stimulating discussions. D.G. and V.D. acknowledge funding from the Israel Science Foundation (grant no. 1395/16). 




\bibliographystyle{mnras}
\bibliography{references} 

\begin{thebibliography}{}
\makeatletter
\relax
\def\mn@urlcharsother{\let\do\@makeother \do\$\do\&\do\#\do\^\do\_\do\%\do\~}
\def\mn@doi{\begingroup\mn@urlcharsother \@ifnextchar [ {\mn@doi@}
  {\mn@doi@[]}}
\def\mn@doi@[#1]#2{\def\@tempa{#1}\ifx\@tempa\@empty \href
  {http://dx.doi.org/#2} {doi:#2}\else \href {http://dx.doi.org/#2} {#1}\fi
  \endgroup}
\def\mn@eprint#1#2{\mn@eprint@#1:#2::\@nil}
\def\mn@eprint@arXiv#1{\href {http://arxiv.org/abs/#1} {{\tt arXiv:#1}}}
\def\mn@eprint@dblp#1{\href {http://dblp.uni-trier.de/rec/bibtex/#1.xml}
  {dblp:#1}}
\def\mn@eprint@#1:#2:#3:#4\@nil{\def\@tempa {#1}\def\@tempb {#2}\def\@tempc
  {#3}\ifx \@tempc \@empty \let \@tempc \@tempb \let \@tempb \@tempa \fi \ifx
  \@tempb \@empty \def\@tempb {arXiv}\fi \@ifundefined
  {mn@eprint@\@tempb}{\@tempb:\@tempc}{\expandafter \expandafter \csname
  mn@eprint@\@tempb\endcsname \expandafter{\@tempc}}}

\bibitem[\protect\citeauthoryear{{Alonso} \& {Ferreira}}{{Alonso} \&
  {Ferreira}}{2015b}]{2015PhRvD..92f3525A}
{Alonso} D.,  {Ferreira} P.~G.,  2015b, \mn@doi [\prd]
  {10.1103/PhysRevD.92.063525}, \href
  {https://ui.adsabs.harvard.edu/abs/2015PhRvD..92f3525A} {92, 063525}

\bibitem[\protect\citeauthoryear{{Alonso} \& {Ferreira}}{{Alonso} \&
  {Ferreira}}{2015a}]{alonso/ferreira:2015}
{Alonso} D.,  {Ferreira} P.~G.,  2015a, \mn@doi [\prd]
  {10.1103/PhysRevD.92.063525}, \href
  {https://ui.adsabs.harvard.edu/abs/2015PhRvD..92f3525A} {92, 063525}

\bibitem[\protect\citeauthoryear{{Ando}, {Benoit-L{\'e}vy}  \&
  {Komatsu}}{{Ando} et~al.}{2018}]{Andoetal:2018}
{Ando} S.,  {Benoit-L{\'e}vy} A.,   {Komatsu} E.,  2018, \mn@doi [\mnras]
  {10.1093/mnras/stx2634}, \href
  {https://ui.adsabs.harvard.edu/abs/2018MNRAS.473.4318A} {473, 4318}

\bibitem[\protect\citeauthoryear{Baldauf, Seljak, Senatore  \&
  Zaldarriaga}{Baldauf et~al.}{2011}]{Baldauf:2011bh}
Baldauf T.,  Seljak U.,  Senatore L.,   Zaldarriaga M.,  2011, \mn@doi [JCAP]
  {10.1088/1475-7516/2011/10/031}, 1110, 031

\bibitem[\protect\citeauthoryear{Baldauf, Seljak, Smith, Hamaus  \&
  Desjacques}{Baldauf et~al.}{2013}]{Baldauf:2013hka}
Baldauf T.,  Seljak U.,  Smith R.~E.,  Hamaus N.,   Desjacques V.,  2013,
  \mn@doi [Phys. Rev.] {10.1103/PhysRevD.88.083507}, D88, 083507

\bibitem[\protect\citeauthoryear{{Bardeen}, {Bond}, {Kaiser}  \&
  {Szalay}}{{Bardeen} et~al.}{1986}]{bbks}
{Bardeen} J.~M.,  {Bond} J.~R.,  {Kaiser} N.,   {Szalay} A.~S.,  1986, \mn@doi
  [\apj] {10.1086/164143}, \href
  {https://ui.adsabs.harvard.edu/abs/1986ApJ...304...15B} {304, 15}

\bibitem[\protect\citeauthoryear{Beltran~Jimenez \& Durrer}{Beltran~Jimenez \&
  Durrer}{2011}]{BeltranJimenez:2010bb}
Beltran~Jimenez J.,  Durrer R.,  2011, \mn@doi [Phys. Rev.]
  {10.1103/PhysRevD.83.103509}, D83, 103509

\bibitem[\protect\citeauthoryear{Biagetti, Lazeyras, Baldauf, Desjacques  \&
  Schmidt}{Biagetti et~al.}{2017}]{Biagetti:2016ywx}
Biagetti M.,  Lazeyras T.,  Baldauf T.,  Desjacques V.,   Schmidt F.,  2017,
  \mn@doi [Mon. Not. Roy. Astron. Soc.] {10.1093/mnras/stx714}, 468, 3277

\bibitem[\protect\citeauthoryear{{Bonvin} \& {Durrer}}{{Bonvin} \&
  {Durrer}}{2011}]{bonvin/durrer:2011}
{Bonvin} C.,  {Durrer} R.,  2011, \mn@doi [\prd] {10.1103/PhysRevD.84.063505},
  \href {https://ui.adsabs.harvard.edu/abs/2011PhRvD..84f3505B} {84, 063505}

\bibitem[\protect\citeauthoryear{{Cai}, {Bernstein}  \& {Sheth}}{{Cai}
  et~al.}{2011}]{cai/etal:2011}
{Cai} Y.-C.,  {Bernstein} G.,   {Sheth} R.~K.,  2011, \mn@doi [\mnras]
  {10.1111/j.1365-2966.2010.17969.x}, \href
  {https://ui.adsabs.harvard.edu/abs/2011MNRAS.412..995C} {412, 995}

\bibitem[\protect\citeauthoryear{Camera, Santos  \& Maartens}{Camera
  et~al.}{2015}]{camera/etal:2015}
Camera S.,  Santos M.~G.,   Maartens R.,  2015, \mn@doi [Monthly Notices of the
  Royal Astronomical Society] {10.1093/mnras/stv040}, 448, 1035

\bibitem[\protect\citeauthoryear{{Casas-Miranda}, {Mo}, {Sheth}  \&
  {Boerner}}{{Casas-Miranda} et~al.}{2002}]{CasasMiranda:2002on}
{Casas-Miranda} R.,  {Mo} H.~J.,  {Sheth} R.~K.,   {Boerner} G.,  2002, \mn@doi
  [\mnras] {10.1046/j.1365-8711.2002.05378.x}, \href
  {http://adsabs.harvard.edu/abs/2002MNRAS.333..730C} {333, 730}

\bibitem[\protect\citeauthoryear{{Castorina}, {Feng}, {Seljak}  \&
  {Villaescusa-Navarro}}{{Castorina} et~al.}{2018}]{castorina/seljak/etal:2018}
{Castorina} E.,  {Feng} Y.,  {Seljak} U.,   {Villaescusa-Navarro} F.,  2018,
  \mn@doi [\prl] {10.1103/PhysRevLett.121.101301}, \href
  {https://ui.adsabs.harvard.edu/abs/2018PhRvL.121j1301C} {121, 101301}

\bibitem[\protect\citeauthoryear{{Catelan}, {Lucchin}, {Matarrese}  \&
  {Porciani}}{{Catelan} et~al.}{1998}]{catelan/etal:1999}
{Catelan} P.,  {Lucchin} F.,  {Matarrese} S.,   {Porciani} C.,  1998, \mn@doi
  [\mnras] {10.1046/j.1365-8711.1998.01455.x}, \href
  {https://ui.adsabs.harvard.edu/abs/1998MNRAS.297..692C} {297, 692}

\bibitem[\protect\citeauthoryear{{Challinor} \& {Lewis}}{{Challinor} \&
  {Lewis}}{2011}]{lewis/challinor:2011}
{Challinor} A.,  {Lewis} A.,  2011, \mn@doi [\prd]
  {10.1103/PhysRevD.84.043516}, \href
  {https://ui.adsabs.harvard.edu/abs/2011PhRvD..84d3516C} {84, 043516}

\bibitem[\protect\citeauthoryear{{Cole} \& {Kaiser}}{{Cole} \&
  {Kaiser}}{1989}]{cole/kaiser:1989}
{Cole} S.,  {Kaiser} N.,  1989, \mnras, \href
  {http://adsabs.harvard.edu/abs/1989MNRAS.237.1127C} {237, 1127}

\bibitem[\protect\citeauthoryear{{Dalal}, {Dor{\'e}}, {Huterer}  \&
  {Shirokov}}{{Dalal} et~al.}{2008}]{2008PhRvD..77l3514D}
{Dalal} N.,  {Dor{\'e}} O.,  {Huterer} D.,   {Shirokov} A.,  2008, \mn@doi
  [\prd] {10.1103/PhysRevD.77.123514}, \href
  {https://ui.adsabs.harvard.edu/abs/2008PhRvD..77l3514D} {77, 123514}

\bibitem[\protect\citeauthoryear{{Desjacques}, {Jeong}  \&
  {Schmidt}}{{Desjacques} et~al.}{2018}]{Desjacques:2016bnm}
{Desjacques} V.,  {Jeong} D.,   {Schmidt} F.,  2018, \mn@doi [\physrep]
  {10.1016/j.physrep.2017.12.002}, \href
  {https://ui.adsabs.harvard.edu/abs/2018PhR...733....1D} {733, 1}

\bibitem[\protect\citeauthoryear{{Dor{\'e}} et~al.,}{{Dor{\'e}}
  et~al.}{2014}]{SPHEREx}
{Dor{\'e}} O.,  et~al., 2014, arXiv e-prints, \href
  {https://ui.adsabs.harvard.edu/abs/2014arXiv1412.4872D} {p. arXiv:1412.4872}

\bibitem[\protect\citeauthoryear{Durrer, Gabrielli, Joyce  \&
  Sylos~Labini}{Durrer et~al.}{2003}]{Durrer:2002zu}
Durrer R.,  Gabrielli A.,  Joyce M.,   Sylos~Labini F.,  2003, \mn@doi
  [Astrophys. J.] {10.1086/374208}, 585, L1

\bibitem[\protect\citeauthoryear{{Feldman}, {Kaiser}  \& {Peacock}}{{Feldman}
  et~al.}{1994}]{FKP}
{Feldman} H.~A.,  {Kaiser} N.,   {Peacock} J.~A.,  1994, \mn@doi [\apj]
  {10.1086/174036}, \href
  {https://ui.adsabs.harvard.edu/abs/1994ApJ...426...23F} {426, 23}

\bibitem[\protect\citeauthoryear{{Fonseca}, {Camera}, {Santos}  \&
  {Maartens}}{{Fonseca} et~al.}{2015}]{2015ApJ...812L..22F}
{Fonseca} J.,  {Camera} S.,  {Santos} M.~G.,   {Maartens} R.,  2015, \mn@doi
  [\apjl] {10.1088/2041-8205/812/2/L22}, \href
  {https://ui.adsabs.harvard.edu/abs/2015ApJ...812L..22F} {812, L22}

\bibitem[\protect\citeauthoryear{{Gaztanaga}, {Bonvin}  \& {Hui}}{{Gaztanaga}
  et~al.}{2017}]{gaztanaga/etal:2017}
{Gaztanaga} E.,  {Bonvin} C.,   {Hui} L.,  2017, \mn@doi [\jcap]
  {10.1088/1475-7516/2017/01/032}, \href
  {https://ui.adsabs.harvard.edu/abs/2017JCAP...01..032G} {2017, 032}

\bibitem[\protect\citeauthoryear{{Ginzburg}, {Desjacques}  \&
  {Chan}}{{Ginzburg} et~al.}{2017}]{2017PhRvD..96h3528G}
{Ginzburg} D.,  {Desjacques} V.,   {Chan} K.~C.,  2017, \mn@doi [\prd]
  {10.1103/PhysRevD.96.083528}, \href
  {https://ui.adsabs.harvard.edu/abs/2017PhRvD..96h3528G} {96, 083528}

\bibitem[\protect\citeauthoryear{{Gomes}, {Camera}, {Jarvis}, {Hale}  \&
  {Fonseca}}{{Gomes} et~al.}{2020}]{2020MNRAS.492.1513G}
{Gomes} Z.,  {Camera} S.,  {Jarvis} M.~J.,  {Hale} C.,   {Fonseca} J.,  2020,
  \mn@doi [\mnras] {10.1093/mnras/stz3581}, \href
  {https://ui.adsabs.harvard.edu/abs/2020MNRAS.492.1513G} {492, 1513}

\bibitem[\protect\citeauthoryear{Hamaus, Seljak, Desjacques, Smith  \&
  Baldauf}{Hamaus et~al.}{2010}]{Hamaus:2010im}
Hamaus N.,  Seljak U.,  Desjacques V.,  Smith R.~E.,   Baldauf T.,  2010,
  \mn@doi [Phys. Rev.] {10.1103/PhysRevD.82.043515}, D82, 043515

\bibitem[\protect\citeauthoryear{{Hamaus}, {Seljak}  \& {Desjacques}}{{Hamaus}
  et~al.}{2011}]{2011PhRvD..84h3509H}
{Hamaus} N.,  {Seljak} U.,   {Desjacques} V.,  2011, \mn@doi [\prd]
  {10.1103/PhysRevD.84.083509}, \href
  {https://ui.adsabs.harvard.edu/abs/2011PhRvD..84h3509H} {84, 083509}

\bibitem[\protect\citeauthoryear{Hamaus, Seljak  \& Desjacques}{Hamaus
  et~al.}{2012}]{Hamaus:2012ap}
Hamaus N.,  Seljak U.,   Desjacques V.,  2012, \mn@doi [Phys. Rev.]
  {10.1103/PhysRevD.86.103513}, D86, 103513

\bibitem[\protect\citeauthoryear{{Hamilton} \& {Culhane}}{{Hamilton} \&
  {Culhane}}{1996}]{hamilton/culhane:1996}
{Hamilton} A.~J.~S.,  {Culhane} M.,  1996, \mn@doi [\mnras]
  {10.1093/mnras/278.1.73}, \href
  {https://ui.adsabs.harvard.edu/abs/1996MNRAS.278...73H} {278, 73}

\bibitem[\protect\citeauthoryear{{Heavens}}{{Heavens}}{2009}]{2009arXiv0906.0664H}
{Heavens} A.,  2009, preprint, \href
  {http://adsabs.harvard.edu/abs/2009arXiv0906.0664H} {} (\mn@eprint {arXiv}
  {0906.0664})

\bibitem[\protect\citeauthoryear{Heavens, Kitching  \& Verde}{Heavens
  et~al.}{2007}]{Heavens_2007}
Heavens A.~F.,  Kitching T.~D.,   Verde L.,  2007, \mn@doi [Monthly Notices of
  the Royal Astronomical Society] {10.1111/j.1365-2966.2007.12134.x}, 380,
  1029–1035

\bibitem[\protect\citeauthoryear{{Hui}, {Gazta{\~n}aga}  \& {Loverde}}{{Hui}
  et~al.}{2007}]{hui/etal:2007}
{Hui} L.,  {Gazta{\~n}aga} E.,   {Loverde} M.,  2007, \mn@doi [\prd]
  {10.1103/PhysRevD.76.103502}, \href
  {https://ui.adsabs.harvard.edu/abs/2007PhRvD..76j3502H} {76, 103502}

\bibitem[\protect\citeauthoryear{{Jeong}, {Schmidt}  \& {Hirata}}{{Jeong}
  et~al.}{2012}]{jeong/schmidt/hirata:2012}
{Jeong} D.,  {Schmidt} F.,   {Hirata} C.~M.,  2012, \mn@doi [\prd]
  {10.1103/PhysRevD.85.023504}, \href
  {https://ui.adsabs.harvard.edu/abs/2012PhRvD..85b3504J} {85, 023504}

\bibitem[\protect\citeauthoryear{{Kaiser}}{{Kaiser}}{1984}]{kaiser:1984}
{Kaiser} N.,  1984, \mn@doi [\apjl] {10.1086/184341}, \href
  {http://adsabs.harvard.edu/abs/1984ApJ...284L...9K} {284, L9}

\bibitem[\protect\citeauthoryear{{Kaiser}}{{Kaiser}}{1987}]{kaiser:1987}
{Kaiser} N.,  1987, \mnras, \href
  {http://adsabs.harvard.edu/abs/1987MNRAS.227....1K} {227, 1}

\bibitem[\protect\citeauthoryear{{Komatsu} et~al.,}{{Komatsu}
  et~al.}{2011}]{2011ApJS..192...18K}
{Komatsu} E.,  et~al., 2011, \mn@doi [\apjs] {10.1088/0067-0049/192/2/18},
  \href {https://ui.adsabs.harvard.edu/abs/2011ApJS..192...18K} {192, 18}

\bibitem[\protect\citeauthoryear{{Matarrese} \& {Verde}}{{Matarrese} \&
  {Verde}}{2008}]{2008ApJ...677L..77M}
{Matarrese} S.,  {Verde} L.,  2008, \mn@doi [\apjl] {10.1086/587840}, \href
  {https://ui.adsabs.harvard.edu/abs/2008ApJ...677L..77M} {677, L77}

\bibitem[\protect\citeauthoryear{{Matsubara}}{{Matsubara}}{2000}]{matsubara:2000}
{Matsubara} T.,  2000, \mn@doi [\apj] {10.1086/308827}, \href
  {https://ui.adsabs.harvard.edu/abs/2000ApJ...535....1M} {535, 1}

\bibitem[\protect\citeauthoryear{{McDonald}}{{McDonald}}{2009}]{mcdonald:2009}
{McDonald} P.,  2009, \mn@doi [\jcap] {10.1088/1475-7516/2009/11/026}, \href
  {https://ui.adsabs.harvard.edu/abs/2009JCAP...11..026M} {2009, 026}

\bibitem[\protect\citeauthoryear{{McDonald} \& {Seljak}}{{McDonald} \&
  {Seljak}}{2009}]{mcdonald/seljak:2009}
{McDonald} P.,  {Seljak} U.,  2009, \mn@doi [\jcap]
  {10.1088/1475-7516/2009/10/007}, \href
  {https://ui.adsabs.harvard.edu/abs/2009JCAP...10..007M} {2009, 007}

\bibitem[\protect\citeauthoryear{Mo \& White}{Mo \& White}{1996}]{Mo:1995cs}
Mo H.~J.,  White S. D.~M.,  1996, \mn@doi [Mon. Not. Roy. Astron. Soc.]
  {10.1093/mnras/282.2.347}, 282, 347

\bibitem[\protect\citeauthoryear{{Moessner}, {Jain}  \& {Villumsen}}{{Moessner}
  et~al.}{1998}]{moessner/etal:1998}
{Moessner} R.,  {Jain} B.,   {Villumsen} J.~V.,  1998, \mn@doi [\mnras]
  {10.1046/j.1365-8711.1998.01225.x}, \href
  {https://ui.adsabs.harvard.edu/abs/1998MNRAS.294..291M} {294, 291}

\bibitem[\protect\citeauthoryear{{Moradinezhad Dizgah} \&
  {Keating}}{{Moradinezhad Dizgah} \& {Keating}}{2019}]{2019ApJ...872..126M}
{Moradinezhad Dizgah} A.,  {Keating} G.~K.,  2019, \mn@doi [\apj]
  {10.3847/1538-4357/aafd36}, \href
  {https://ui.adsabs.harvard.edu/abs/2019ApJ...872..126M} {872, 126}

\bibitem[\protect\citeauthoryear{{Moradinezhad Dizgah}, {Keating}  \&
  {Fialkov}}{{Moradinezhad Dizgah} et~al.}{2019}]{2019ApJ...870L...4M}
{Moradinezhad Dizgah} A.,  {Keating} G.~K.,   {Fialkov} A.,  2019, \mn@doi
  [\apjl] {10.3847/2041-8213/aaf813}, \href
  {https://ui.adsabs.harvard.edu/abs/2019ApJ...870L...4M} {870, L4}

\bibitem[\protect\citeauthoryear{Paech, Hamaus, Hoyle, Costanzi, Giannantonio,
  Hagstotz, Sauerwein  \& Weller}{Paech et~al.}{2017}]{Paech:2016hod}
Paech K.,  Hamaus N.,  Hoyle B.,  Costanzi M.,  Giannantonio T.,  Hagstotz S.,
  Sauerwein G.,   Weller J.,  2017, \mn@doi [Monthly Notices of the Royal
  Astronomical Society] {10.1093/mnras/stx1354}, 470, 2566–2577

\bibitem[\protect\citeauthoryear{Raccanelli, Dore  \& Dalal}{Raccanelli
  et~al.}{2015}]{Raccanelli:2014awa}
Raccanelli A.,  Dore O.,   Dalal N.,  2015, \mn@doi [JCAP]
  {10.1088/1475-7516/2015/08/034}, 1508, 034

\bibitem[\protect\citeauthoryear{{Seljak}}{{Seljak}}{2009}]{2009PhRvL.102b1302S}
{Seljak} U.,  2009, \mn@doi [\prl] {10.1103/PhysRevLett.102.021302}, \href
  {https://ui.adsabs.harvard.edu/abs/2009PhRvL.102b1302S} {102, 021302}

\bibitem[\protect\citeauthoryear{Seljak, Hamaus  \& Desjacques}{Seljak
  et~al.}{2009}]{Seljak:2009af}
Seljak U.,  Hamaus N.,   Desjacques V.,  2009, \mn@doi [Phys. Rev. Lett.]
  {10.1103/PhysRevLett.103.091303}, 103, 091303

\bibitem[\protect\citeauthoryear{Sheth \& Lemson}{Sheth \&
  Lemson}{1999}]{Sheth:1998xe}
Sheth R.~K.,  Lemson G.,  1999, \mn@doi [Mon. Not. Roy. Astron. Soc.]
  {10.1046/j.1365-8711.1999.02378.x}, 304, 767

\bibitem[\protect\citeauthoryear{{Sheth} \& {Tormen}}{{Sheth} \&
  {Tormen}}{1999}]{sheth/tormen:1999}
{Sheth} R.~K.,  {Tormen} G.,  1999, \mn@doi [\mnras]
  {10.1046/j.1365-8711.1999.02692.x}, \href
  {http://adsabs.harvard.edu/abs/1999MNRAS.308..119S} {308, 119}

\bibitem[\protect\citeauthoryear{{Slosar}}{{Slosar}}{2009}]{slosar:2009}
{Slosar} A.,  2009, \mn@doi [\jcap] {10.1088/1475-7516/2009/03/004}, \href
  {https://ui.adsabs.harvard.edu/abs/2009JCAP...03..004S} {2009, 004}

\bibitem[\protect\citeauthoryear{{Slosar}, {Hirata}, {Seljak}, {Ho}  \&
  {Padmanabhan}}{{Slosar} et~al.}{2008}]{2008JCAP...08..031S}
{Slosar} A.,  {Hirata} C.,  {Seljak} U.,  {Ho} S.,   {Padmanabhan} N.,  2008,
  \mn@doi [\jcap] {10.1088/1475-7516/2008/08/031}, \href
  {https://ui.adsabs.harvard.edu/abs/2008JCAP...08..031S} {2008, 031}

\bibitem[\protect\citeauthoryear{Smith, Scoccimarro  \& Sheth}{Smith
  et~al.}{2007}]{Smith:2006ne}
Smith R.~E.,  Scoccimarro R.,   Sheth R.~K.,  2007, \mn@doi [Phys. Rev.]
  {10.1103/PhysRevD.75.063512}, D75, 063512

\bibitem[\protect\citeauthoryear{{White}, {Song}  \& {Percival}}{{White}
  et~al.}{2009}]{white/etal:2009}
{White} M.,  {Song} Y.-S.,   {Percival} W.~J.,  2009, \mn@doi [\mnras]
  {10.1111/j.1365-2966.2008.14379.x}, \href
  {https://ui.adsabs.harvard.edu/abs/2009MNRAS.397.1348W} {397, 1348}

\bibitem[\protect\citeauthoryear{{Yamauchi}, {Yokoyama}  \&
  {Takahashi}}{{Yamauchi} et~al.}{2017}]{yamauchi/yokoyama/etal:2017}
{Yamauchi} D.,  {Yokoyama} S.,   {Takahashi} K.,  2017, \mn@doi [\prd]
  {10.1103/PhysRevD.95.063530}, \href
  {https://ui.adsabs.harvard.edu/abs/2017PhRvD..95f3530Y} {95, 063530}

\bibitem[\protect\citeauthoryear{{Yoo}}{{Yoo}}{2010}]{2010PhRvD..82h3508Y}
{Yoo} J.,  2010, \mn@doi [\prd] {10.1103/PhysRevD.82.083508}, \href
  {https://ui.adsabs.harvard.edu/abs/2010PhRvD..82h3508Y} {82, 083508}

\bibitem[\protect\citeauthoryear{{Yoo}, {Fitzpatrick}  \& {Zaldarriaga}}{{Yoo}
  et~al.}{2009}]{2009PhRvD..80h3514Y}
{Yoo} J.,  {Fitzpatrick} A.~L.,   {Zaldarriaga} M.,  2009, \mn@doi [\prd]
  {10.1103/PhysRevD.80.083514}, \href
  {https://ui.adsabs.harvard.edu/abs/2009PhRvD..80h3514Y} {80, 083514}

\bibitem[\protect\citeauthoryear{Yoo, Hamaus, Seljak  \& Zaldarriaga}{Yoo
  et~al.}{2012}]{Yoo:2012se}
Yoo J.,  Hamaus N.,  Seljak U.,   Zaldarriaga M.,  2012, \mn@doi [Phys. Rev.]
  {10.1103/PhysRevD.86.063514}, D86, 063514

\bibitem[\protect\citeauthoryear{de Putter \& Dor\'e}{de~Putter \&
  Dor\'e}{2017a}]{PhysRevD.95.123513}
de Putter R.,  Dor\'e O.,  2017a, \mn@doi [Phys. Rev. D]
  {10.1103/PhysRevD.95.123513}, 95, 123513

\bibitem[\protect\citeauthoryear{de Putter \& Dor\'e}{de~Putter \&
  Dor\'e}{2017b}]{dePutter:2014lna}
de Putter R.,  Dor\'e O.,  2017b, \mn@doi [Phys. Rev.]
  {10.1103/PhysRevD.95.123513}, D95, 123513

\makeatother
\end{thebibliography}




\onecolumn

\appendix

\section{The continuous limit}

Here we show the formulae we used in order to calculate the Fisher matrix for the approximation of infinite number of equal density bins. We follow the derivation of Appendix D in \cite{2011PhRvD..84h3509H}. However, unlike the above reference, we assume that the shot noise is affected only by the "bare" linear bias (in this appendix we define it as $\boldsymbol{b_0}$), rather than the effective bias which contains the corrections due to light propagation (Kaiser effect and general relativistic effects) and corrections due to $f_{\text{\text{NL}}}$ (in this appendix defined as $\boldsymbol{b_c}$).
We recall that according to the halo model the shot noise is $\boldsymbol{\mathcal{E}}=\boldsymbol{\mathcal{A}}-\boldsymbol{\mathcal{M}} \boldsymbol{b_0}^T$ where 
$\boldsymbol{\mathcal{A}}=\bar{n}^{-1}\boldsymbol{I}-\boldsymbol{b_0}\boldsymbol{\mathcal{M}}^T$ and $\boldsymbol{{\mathcal{M}}}=\boldsymbol{M}/{\rho_m}-\boldsymbol{b_0} \left< nM^2 \right> /2 \rho_m^2$ and $\boldsymbol{M}$ is a vector containing the mean halo mass of each bin. And using the Sherman-Morrison formula (see \cite{2011PhRvD..84h3509H}) 
\begin{align}
\boldsymbol{\mathcal{E}}^{-1}&=\boldsymbol{\mathcal{A}}^{-1}+\frac{\boldsymbol{\mathcal{A}}^{-1} \boldsymbol{\mathcal{M}} \boldsymbol{b_0}^T   \boldsymbol{\mathcal{A}}^{-1} }{1-\boldsymbol{b_0}^T \boldsymbol{\mathcal{A}}^{-1} \boldsymbol{\mathcal{M}}} \\ \nonumber
\boldsymbol{\mathcal{A}}^{-1}&=\bar{n} \boldsymbol{I} + \frac{\boldsymbol{b_0} \boldsymbol{\mathcal{M}}^T \bar{n} }{\bar{n}^{-1}-\boldsymbol{\mathcal{M}}^T \boldsymbol{b_0}}
\end{align}
where $\bar{n}$ is the density of one bin.
Therefore
\begin{align}
\boldsymbol{b_c}^\dagger \boldsymbol{\mathcal{A}}^{-1} \boldsymbol{b_c} &= \frac{\boldsymbol{b_c}^\dagger \boldsymbol{b_c} + \bar{n} (\boldsymbol{b_c}^{\dagger} \boldsymbol{b_0} \boldsymbol{\mathcal{M}}^T \boldsymbol{b_c} - \boldsymbol{b_c}^{\dagger} \boldsymbol{b_c} \boldsymbol{\mathcal{M}}^T \boldsymbol{b_0})}{n^{-1}-\boldsymbol{\mathcal{M}}^T \boldsymbol{b_0}} \\ \nonumber
\boldsymbol{b_0}^\dagger \boldsymbol{\mathcal{A}}^{-1} \boldsymbol{b_c} &= \frac{\boldsymbol{b_0}^\dagger \boldsymbol{b_c} + \bar{n} (\boldsymbol{b_0}^{\dagger} \boldsymbol{b_0} \boldsymbol{\mathcal{M}}^T \boldsymbol{b_c} - \boldsymbol{b_0}^{\dagger} \boldsymbol{b_c} \boldsymbol{\mathcal{M}}^T \boldsymbol{b_0})}{n^{-1}-\boldsymbol{\mathcal{M}}^T \boldsymbol{b_0}} \\ \nonumber
\boldsymbol{b_c}^\dagger \boldsymbol{\mathcal{A}}^{-1} \boldsymbol{b_0} &= \frac{\boldsymbol{b_c}^\dagger \boldsymbol{b_0} + \bar{n} (\boldsymbol{b_c}^{\dagger} \boldsymbol{b_0} \boldsymbol{\mathcal{M}}^T \boldsymbol{b_0} - \boldsymbol{b_c}^{\dagger} \boldsymbol{b_0} \boldsymbol{\mathcal{M}}^T \boldsymbol{b_0})}{n^{-1}-\boldsymbol{\mathcal{M}}^T \boldsymbol{b_0}} \\ \nonumber
\boldsymbol{b_c}^\dagger \boldsymbol{\mathcal{A}}^{-1} \boldsymbol{\mathcal{M}} &= \bar{n} \frac{\boldsymbol{b_c}^\dagger \boldsymbol{\mathcal{M}} \left(\bar{n}^{-1}-\boldsymbol{\mathcal{M}}^T \boldsymbol{b_0}\right)+ \boldsymbol{b_c}^T \boldsymbol{b_0} \boldsymbol{\mathcal{M}}^T \boldsymbol{\mathcal{M}} }{n^{-1}-\boldsymbol{\mathcal{M}}^T \boldsymbol{b_0}}  \\ \nonumber
\boldsymbol{b_0}^\dagger \boldsymbol{\mathcal{A}}^{-1} \boldsymbol{\mathcal{M}} &= \bar{n} \frac{\boldsymbol{b_0}^\dagger \boldsymbol{\mathcal{M}} \left(\bar{n}^{-1}-\boldsymbol{\mathcal{M}}^T \boldsymbol{b_0}\right)+ \boldsymbol{b_0}^T \boldsymbol{b_0} \boldsymbol{\mathcal{M}}^T \boldsymbol{\mathcal{M}} }{n^{-1}-\boldsymbol{\mathcal{M}}^T \boldsymbol{b_0}}  \\ \nonumber
\boldsymbol{b_0}^\dagger \boldsymbol{\mathcal{A}}^{-1} \boldsymbol{b_0} &= \frac{\boldsymbol{b_0}^\dagger \boldsymbol{b_0}}{n^{-1}-\boldsymbol{\mathcal{M}}^T\boldsymbol{b_0}}
\end{align}
The terms which contain the derivatives of $\boldsymbol{b_c}$ with the respect to one of the parameters are calculated simply by replacing one (or both)  of $\boldsymbol{b_c}$ vectors by $\boldsymbol{b_{c_i}}$ where $\boldsymbol{b_{c_i}}$ is the derivative of the $\boldsymbol{b_c}$ vector with the respect to parameter $i$.
Therefore, similarly to \cite{2011PhRvD..84h3509H} for $\alpha$, $\beta$ and $\gamma$ we get:
\begin{align}
&\alpha=\boldsymbol{b_c}^\dagger \boldsymbol{\mathcal{E}}^{-1} \boldsymbol{b_c} P(k) =\\ \nonumber
&\frac{\bar{n}^{-1}P(k)}{\lambda_+ \lambda_-}\left[\left(\boldsymbol{b_c}^{\dagger}\boldsymbol{b_c}+\bar{n}\left(\boldsymbol{b_c}^\dagger\boldsymbol{b_0}\boldsymbol{\mathcal{M}}^T\boldsymbol{b_c}-\boldsymbol{b_c}^\dagger\boldsymbol{b_c}\boldsymbol{\mathcal{M}}^T\boldsymbol{b_0}\right)\right)\left(1-\boldsymbol{b_0}^\dagger\boldsymbol{\mathcal{A}}^{-1}\boldsymbol{\mathcal{M}}\right)+\left(\bar{n}^{-1}-\boldsymbol{\mathcal{M}}^T\boldsymbol{b_0}\right)\left(\boldsymbol{b_c}^\dagger \boldsymbol{\mathcal{A}}^{-1} \boldsymbol{\mathcal{M}} \boldsymbol{b_0}^\dagger \boldsymbol{\mathcal{A}}^{-1} \boldsymbol{b_c}\right)\right]
\end{align}
\begin{align}
&\beta=\boldsymbol{b_c}^\dagger \boldsymbol{\mathcal{E}}^{-1} \boldsymbol{b_{c_i}} P(k) = \\ \nonumber &\frac{\bar{n}^{-1}P(k)}{\lambda_+ \lambda_-}\left[\left(\boldsymbol{b_c}^{\dagger}\boldsymbol{b_{c_i}}+\bar{n}\left(\boldsymbol{b_c}^\dagger\boldsymbol{b_0}\boldsymbol{\mathcal{M}}^T\boldsymbol{b_{c_i}}-\boldsymbol{b_c}^\dagger\boldsymbol{b_{c_i}}\boldsymbol{\mathcal{M}}^T\boldsymbol{b_0}\right)\right)\left(1-\boldsymbol{b_0}^\dagger\boldsymbol{\mathcal{A}}^{-1}\boldsymbol{\mathcal{M}}\right)+\left(\bar{n}^{-1}-\boldsymbol{\mathcal{M}}^T\boldsymbol{b_0}\right)\left(\boldsymbol{b_c}^\dagger \boldsymbol{\mathcal{A}}^{-1} \boldsymbol{\mathcal{M}} \boldsymbol{b_0}^\dagger \boldsymbol{\mathcal{A}}^{-1} \boldsymbol{b_{c_i}}\right)\right]
\end{align}
\begin{align}
&\gamma=\boldsymbol{b_{c_i}}^\dagger \boldsymbol{\mathcal{E}}^{-1} \boldsymbol{b_{c_i}} P(k) = \\ \nonumber
&\frac{\bar{n}^{-1}P(k)}{\lambda_+ \lambda_-}\left[\left(\boldsymbol{b_{c_i}}^{\dagger}\boldsymbol{b_{c_i}}+\bar{n}\left(\boldsymbol{b_{c_i}}^\dagger\boldsymbol{b_0}\boldsymbol{\mathcal{M}}^T\boldsymbol{b_{c_i}}-\boldsymbol{b_{c_i}}^\dagger\boldsymbol{b_{c_i}}\boldsymbol{\mathcal{M}}^T\boldsymbol{b_0}\right)\right)\left(1-\boldsymbol{b_0}^\dagger\boldsymbol{\mathcal{A}}^{-1}\boldsymbol{\mathcal{M}}\right)+\left(\bar{n}^{-1}-\boldsymbol{\mathcal{M}}^T\boldsymbol{b_0}\right)\left(\boldsymbol{b_{c_i}}^\dagger \boldsymbol{\mathcal{A}}^{-1} \boldsymbol{\mathcal{M}} \boldsymbol{b_0}^\dagger \boldsymbol{\mathcal{A}}^{-1} \boldsymbol{b_{c_i}}\right)\right]
\end{align}
where, similarly to \cite{2011PhRvD..84h3509H} we define
\begin{align}
\lambda_+ \lambda_- =\left(\bar{n}^{-1}-\boldsymbol{\mathcal{M}}^T\boldsymbol{b_0}\right)^2-\boldsymbol{b_0}^T \boldsymbol{b_0} \boldsymbol{\mathcal{M}}^T \boldsymbol{\mathcal{M}}
\end{align}
From now, each term of the form $\boldsymbol{x}^T\boldsymbol{y}$ or $\boldsymbol{x}^\dagger\boldsymbol{y}$ is calculated using equations 55-56 in \cite{2011PhRvD..84h3509H}
\begin{align}
\boldsymbol{x}^T\boldsymbol{y}\rightarrow\frac{N}{\bar{n}_{tot}}\int_{M_{\text{min}}}^{M_{\text{max}}}n(M)x(M)y(M)dM \\ \nonumber
\boldsymbol{x}^\dagger\boldsymbol{y}\rightarrow\frac{N}{\bar{n}_{tot}}\int_{M_{\text{min}}}^{M_{\text{max}}}n(M)x^*(M)y(M)dM 
\end{align}
where $N$ is the number of bins and $\bar{n}_{tot}$ is the total halo density, i.e.
\begin{align}
\bar{n}_{tot}=\int_{M_{\text{min}}}^{M_{\text{max}}}n(M)dM
\end{align}
Note that the final expressions for  $\alpha$, $\beta$ and $\gamma$ do not depend on $N$.

\label{lastpage}
\end{document}